# Intensity of the Spontaneous Radiation of Electrons Moving along a Standing Laser Wave

K.S. Badikyan *

New Mexico University, Albuquerque, NM, USA

* badikyan.kar@gmail.com

The radiation of harmonics during the interaction of nonrelativistic electrons with a powerful standing laser wave is studied. Expressions for the harmonic frequencies and the corresponding intensities of the spontaneous radiation are obtained with the aid of the quasiclassical approximation.

1. INTRODUCTION

The interaction of an electron with a strong traveling electromagnetic wave was first described in [1]. Multiphoton Compton scattering in the field of a traveling plane wave was studied in [2]. The problem of the elastic scattering of nonrelativistic electrons in the field of a standing electromagnetic wave was first studied by Kapitza and Dirac [3]. The experimental observation of this effect was described in [4]. The Kapitza-Dirac effect in the field of a strong standing electromagnetic wave was investigated in [5]. The modulation and acceleration of a beam of nonrelativistic electrons in the field of two counter propagating waves were studied in [6-10]. In [7] it was shown, specifically, that the motion of an electron, averaged over fast oscillations, in the field of a standing wave can be described with the aid of a ponderomotive potential (Gaponov-Miller potential). The channeling of nonrelativistic electron beam in an intense standing light wave was studied by Fedorov, Oganesyan and Prokhorov [11-13]. The other schemes of Free Electron Lasers are presented in [15-65].

The capabilities of the laser technology existing at that time were, however, inadequate for experimentally checking the results obtained in [10], and the frequencies calculated [10] were limited to the far-IR range. The electromagnetic radiation intensities that can now be achieved for picoseconds and subpicosecond laser pulses with intensity $\sim 10^{16} W/cm^2$ and higher at the focus make it possible to observe emission of near-IR and optical range photons by nonrelativistic electrons.

In the present paper, expressions for the harmonic frequencies and the corresponding intensities of the spontaneous radiation are obtained with the aid of the quasiclassical approximation.

## 2. FORMULATION OF THE PROBLEM. BASIC EQUATIONS

Let us consider the behavior of a nonrelativistic electron in the field of a linearly polarized electromagnetic standing wave. Let the field of the wave be given by the classical vector potential

$$\mathbf{A}_1(z,t) = 2A_{01}\mathbf{e}_1 \cos(\omega_1 t)\sin(k_1 z), \qquad (1)$$

where $A_{01}$ and $k_1(\omega, \pm \mathbf{k}_1)$ are, respectively, the amplitude and 4-momentum of waves which propagate in opposite directions along the z axis and form a standing wave, and $\mathbf{e}_1 = \mathbf{e}_x$ is the unit polarization vector directed along the *x* axis.

We assume that the initial electron momentum **p** makes a small angle with the direction of the standing wave (the z axis), so that the longitudinal component of the momentum is much greater than the transverse component, $p_\parallel \gg p_\perp$. Generally speaking, an electron can be transmitted at an arbitrary angle relative to the direction of the wave. As will be shown below, however, the intensity of the harmonic radiation is proportional to the squared longitudinal length of the wave-electron interaction region, and for this reason the geometry in which the particles are directed along the wave is preferred.

In this geometry, since the ratio $e(\mathbf{A}_1 \cdot \mathbf{p})/(eA_1)^2 \ll 1$ the term $((eA_1)^2$ in the operator ($\hbar = c = 1$)

$$\widehat{V} = \frac{e(\mathbf{A}_1 \cdot \mathbf{p})}{m_e} + \frac{(eA_1)^2}{2m_e}, \qquad (2)$$

is responsible for the interaction of the electron with the wave.

The Schrodinger equation for a particle in the field of a standing wave has the form

$$i\dot{\Psi} = -\frac{1}{2m_e}\Delta\Psi + V_0 \sin^2(k_1 z)\Psi, \qquad (3)$$

where $V_0 = (eA_{01})^2/m_e$ is the amplitude of the effective potential (Gaponov-Miller potential).

The solution of Eq. (3) is sought as a product of functions of the transverse coordinates *x* and *y* and the longitudinal coordinate *z* of the electron:

$$\Psi(\mathbf{r},t) = \exp[-i(\varepsilon t - \mathbf{p}_\perp \cdot \boldsymbol{\rho})]\psi(z), \quad (4)$$

where $\varepsilon = p^2/2m_e$ is the kinetic energy of the electron in the case when the field of the wave is switched off adiabatically and $\boldsymbol{\rho}(x,y)$ is the radius vector of the electron in the *xy* plane.

Substituting the expression (4) into Eq. (3) gives an equation for $\psi(z)$,

$$\varepsilon_\| \psi(z) = -\frac{1}{2m_e}\psi''(z) + V_0 \sin^2(k_1 z)\psi(z), \quad (5)$$

where $\varepsilon_\| = \varepsilon - \varepsilon_\perp$ and $\varepsilon_\perp = p_\perp^2/2m_e$ are the components of the electron energy associated with unperturbed motion of the electron parallel and transverse to the wave, and $\varepsilon_\| \approx \varepsilon$.

The solution of Eq. (5) obtained in the quasiclassical approximation has the form

$$\psi(z) = \exp\left[i\int \sqrt{2m_e[\varepsilon_\| - V(z)]}\,dz\right]. \quad (6)$$

The quasiclassicity condition $|\partial \lambda_{D\|}/\partial z| \ll 1$, where $\lambda_{D\|}$ is the de Broglie wavelength associated with the above-barrier motion of the electron along the wave, is given in terms of our problem by the inequality

$$\frac{m}{(1-m)^{3/2}} \frac{\omega_1}{(m_e \varepsilon_\|)^{1/2}} \ll 1,$$

and for initial electron energy $\varepsilon \gg \omega$ where $\omega$ is the frequency of the emitted photon, it imposes an upper limit on how close the parameter *m* can approach 1. Since the ratio $\omega_1/(m_e \varepsilon_\|)^{1/2} \sim 10^{-5}$ is small, the condition for quasiclassical electron motion is consistent with the inequality $1-m \ll 1$ (for $\varepsilon_\| \approx 10 KeV$).

The field of the emitted wave with the vector potential

$$\mathbf{A}_2 = (A_{02}/2)[\mathbf{e}\exp[i(\omega_1 t - \mathbf{k}\cdot\mathbf{r})] + \text{c.c.}] \quad (7)$$

($A_{02}$ and **e** are, respectively, the amplitude and polarization unit vector) is assumed to be weak and is taken into account by means of perturbation theory.

To first order in the field (7) the amplitude for the transition of an electron from the initial state $(\varepsilon, \mathbf{p})$ into the final state $(\varepsilon', \mathbf{p}')$ with the emission of a photon $(\omega, \mathbf{k})$ is determined

by the expression

$$A_{fi}(t) = -i\frac{eA_{02}}{2m_e V}\int_0^t\int \exp[i(\varepsilon'+\omega-\varepsilon)t_1 - (\mathbf{p}'_\perp + \mathbf{k}_\perp + \mathbf{p}_\perp)\psi_f^*(z)\left(\mathbf{e}_\perp \mathbf{p}_\perp - ie_z\frac{\partial}{\partial z}\right)$$
$$\times \exp(-ik_z z)\psi_i(z)d\mathbf{\rho} dz dt_1, \quad (8)$$

where $\psi_i(z)$ and $\psi_f(z)$ are, respectively, the initial and final wave functions (6) and V is the normalization volume of the particle.

Assuming that the relative change in the longitudinal component of the electron momentum during the emission process is small, $|\Delta p_\parallel / p_\parallel| \sim \omega/[(1-m)\varepsilon_\parallel] \ll 1$, after integrating in Eq. (8) over the time and the transverse coordinates we obtain

$$A_{fi}(t) = -i\frac{eA_{02}}{2m_e k_1}\frac{\exp[i(\varepsilon'+\omega-\varepsilon)t - 1}{i(\varepsilon'+\omega-\varepsilon)}\left[\frac{2J_1(u)}{u}\right]\frac{1}{l}$$
$$\times \left\{(\mathbf{e}_\perp \mathbf{p}_\perp)\int_{-\pi N/2}^{\pi N/2} \exp[-i\Delta F(\xi|m)\right.$$
$$+(k_z/\Delta p_\parallel)\xi]d\xi + (e_z p_\parallel)\int_{-\pi N/2}^{\pi N/2}(1-m\sin^2\xi)^{1/2}\exp[-i\Delta F(\xi|m)$$
$$\left. + (k_z/\Delta p_\parallel)\xi]d\xi\right\}. \quad (9)$$

In Eq. (9) we introduced the notation

$$u = |\Delta\mathbf{p}_\perp + \mathbf{k}_\perp|\rho_0 \sin\theta, \quad (10)$$

where $\theta$ is the angle made by the vector $\Delta\mathbf{p}+\mathbf{k}$ with the z axis; $\Delta p_\parallel = p'_\parallel - p_\parallel$ and $\Delta\mathbf{p}_\perp = \mathbf{p}'_\perp - \mathbf{p}_\perp$ are, respectively, the changes in the longitudinal and transverse components of the electron momentum; $\Delta = \Delta p_\parallel / k_\parallel$; $N = 2l/\lambda_1$ is the number of standing waves which fit within the electron-wave interaction region; $l = v_\parallel \tau$ is the distance traversed by an electron along the standing wave during the pulse time $\tau$; $\lambda_1 = 2\pi/k_1$;

$$F(\xi|m) = \int_0^\xi (1-m\sin^2 x)^{-1/2}dx$$

is an elliptic integral of the first kind [14]; and, $J_1(u)$ is a Bessel function.

In Eq. (9) it is assumed that $V = \pi\rho_0^2 l$, where $\rho_0$ is the radius of the focus at the center of the standing wave (in the plane $z = 0$).

Squaring the expression (9) and using the well-known representations for a delta function ($\omega_1 \tau \gg 1, \rho_0 \gg \lambda$, where $\lambda$ is the spontaneous-emission wavelength), we obtain for the probability of a transition per unit time into the partial final state

$$\frac{|A_{fi}(t)|}{t} = \left(\frac{eA_{02}}{2m_e}\right)^2 \frac{2\lambda_1^2}{\rho_0^2} \delta(\varepsilon' + \omega - \varepsilon)\delta^{(2)}(\Delta \mathbf{p}_\perp + \mathbf{k}_\perp) \frac{1}{l^2}|(\mathbf{e}_\perp \mathbf{p}_\perp)I_1 + (e_z p_\parallel)I_2|^2, \quad (11)$$

where

$$I_1 = \int_{-\pi N/2}^{\pi N/2} \exp\{i[|\Delta|F(\xi|m) - \beta\xi]\} d\xi,$$

$$I_2 = \int_{-\pi N/2}^{\pi N/2} (1 - m\sin^2\xi)^{1/2} \exp\{i[|\Delta|F(\xi|m) - \beta\xi]\} d\xi, \quad (12)$$

$$\beta \equiv k_z / k_1.$$

Integrating over the transverse momentum of the scattered electron, we obtain from Eq. (11) the following expression for the spectral-angular density of the intensity of the spontaneous radiation:

$$\frac{d^2 I}{d\omega d\Omega_\mathbf{k}} = \frac{\alpha \lambda_1^2 \omega^2}{\pi^2 (2m_e)^2} \delta(\varepsilon' + \omega - \varepsilon)\delta^{(2)}(\Delta \mathbf{p}_\perp + \mathbf{k}_\perp) \frac{1}{l} |(\mathbf{e}_\perp \mathbf{p}_\perp)I_1 + (e_z p_\parallel)I_2|^2 dp_\parallel, \quad (13)$$

where $\alpha = e^2/\hbar c$ is the fine structure constant.

From the law of conservation of the transverse component of the momentum of the system $\mathbf{p}'_\perp + \mathbf{k}_\perp - \mathbf{p}_\perp = 0$ it is easy to express the final energy $\varepsilon'$ of the electron in terms of the parameters of its initial state and the characteristics of the emitted photon:

$$\varepsilon' = (p'_\parallel + p'_\perp)^2 / 2m_e = [p_\perp^2 - 2(\mathbf{p}_\perp \mathbf{k}_\perp) + k_\perp^2 + p'^2_\parallel]/2m_e. \quad (14)$$

Using a well-known property of delta functions, with the aid of Eq. (14) we represent $\delta(\varepsilon' + \omega - \varepsilon)$ in the form

$$\delta(\varepsilon' + \omega - \varepsilon) = \frac{m_e}{p_0}[\delta(p'_\parallel - p_0) + \delta(p'_\parallel + p_0)], \quad (15)$$

where

$$p_0 = (p_\parallel^2 + 2(\mathbf{p}_\perp \mathbf{k}_\perp) - 2m_e \omega - k_\perp^2)^{1/2} \quad (16)$$

Substituting the expression (15) into Eq. (13) and integrating over $dp'_\parallel$, we obtain

$$\frac{d^2I}{d\omega d\Omega_{\mathbf{k}}} = \frac{2\alpha\lambda_1^2\omega^2}{(2\pi)^4(2m_e)^{3/2}\varepsilon^{1/2}}\frac{1}{l}\left|(\mathbf{e}_\perp \mathbf{p}_\perp)I_1 + (e_z p_\parallel)I_2\right|^2 dp_\parallel, \tag{17}$$

where the approximate equality $p_0 \approx p_\parallel \approx (2m_e\varepsilon)^{1/2}$ was used in the denominator.

The expression (17) is general, and further calculations depend on the value taken for the parameter m.